\documentstyle[epsfig,12pt,a4p,subfigure]{article}

\newcommand{\be}{\begin{equation}}
\newcommand{\ee}{\end{equation}}
\newcommand{\bea}{\begin{eqnarray}}
\newcommand{\eea}{\end{eqnarray}}
\newcommand{\bean}{\begin{eqnarray*}}
\newcommand{\eean}{\end{eqnarray*}}

\newcommand{\pom}{$I\hspace{-1.6mm}P$}
\newcommand{\pome}{I\hspace{-1.6mm}P}

\newcommand{\gapproxeq}{\lower
.7ex\hbox{$\;\stackrel{\textstyle >}{\sim}\;$}}
\newcommand{\lapproxeq}{\lower
.7ex\hbox{$\;\stackrel{\textstyle <}{\sim}\;$}}

\begin{document}
\begin{titlepage}
\begin{tabbing}
wwwwwwwwwwwwwwwright hand corner using tabbing so it looks neat and in \= \kill
\> {OUTP-01-29P}   \\
\> {11 June 2001}
\end{tabbing}
\baselineskip=18pt
\vskip 0.7in
\begin{center}
{\bf \LARGE Large Isospin mixing
 in $\phi$ radiative decay and the spatial size of the
$f_0(980)- a_0(980)$ mesons}\\
\vspace*{0.9in}
{\large Frank E. Close}\footnote{\tt{e-mail: F.Close1@physics.ox.ac.uk}} \\
\vspace{.1in}
{\it Theoretical Physics}\\
{\it Oxford University, 1 Keble Rd, Oxford OX1 3NP, UK}\\
\vspace{0.1in}
{\large Andrew Kirk}\footnote{\tt{e-mail: ak@hep.ph.bham.ac.uk}} \\
{\it School of Physics and Astronomy}\\
{\it Birmingham University}\\
\end{center}
\begin{abstract}
The measured rate for $\phi \to \gamma f_0(980)$ appears to
be larger than allowed on rather general grounds.
We show that mixing between the $f_0(980)$ and $a_0(980)$,
due to their dynamical
interaction with the nearby $K\bar{K}$ thresholds,
radically affects some existing
predictions of their production in $\phi$ radiative decay.
We predict that $\Gamma(\phi \to \gamma
f_0)/\Gamma(\phi \to \gamma a_0) \sim 3$;
that $\sum (b.r.(\phi \to \gamma
f_0) + b.r. (\phi \to \gamma a_0)) < 5 \times 10^{-4} $ with
probable individual
branching ratios $\sim 2 \times 10^{-4}$
and $0.7 \times 10^{-4}$ respectively.
\end{abstract}
\end{titlepage}
\setcounter{page}{2}
\par

The radiative decay of the $\phi$ to the enigmatic scalar mesons $f_0(980)$ and
$a_0(980)$ has long been recognised as a potential route towards disentangling
their nature. Particular interest has focussed on the likelihood that these
states
contain significant non-$q\bar{q}$ content, specifically being $(u\bar{u} \pm
d\bar{d})
s\bar{s}$. Such a configuration could either be confined within $R \sim
\Lambda_{QCD}^{-1}$ as a
``four-quark" state\cite{jaffe,achasov3}, or more spatially dispersed into two
identifiable colour-singlets: the $K\bar{K}$ molecule
scenario\cite{iswein,cik}.
In more sophisticated pictures these states could be seeded by an
underlying $q\bar{q}$ or compact $qq\bar{q}\bar{q}$
component, influenced by the S-wave $K\bar{K}$ and related thresholds in this
mass region\cite{jaffe2,tornqv}. Furthermore, significant isospin mixing
effects
are anticipated (and seen) in the neutral $f_0 -a_0^0$ states due to the
nearness of the $K^+K^-$ and $K^0\bar{K^0}$
thresholds\cite{achasov1,ck1,speth}.
In this letter we note that such isospin mixing effects could considerably
alter
some predictions in the literature for $\Gamma(\phi \to f_0(980)\gamma )$ and
$\Gamma(\phi \to  a_0(980)\gamma)$.

The magnitudes of these widths are predicted to be rather
sensitive to the fundamental structures
of the $f_0$ and $a_0$, and as such potentially discriminate amongst them.
For example, if $f_0(980) \equiv s\bar{s}$ and the dominant dynamics is
the ``direct" quark transition
$\phi(s\bar{s}) \to \gamma 0^{++}(s\bar{s})$, then
the predicted $b.r.(\phi \to \gamma f_0)
\sim 10^{-5}$, the rate to $\phi \to \gamma a_0(q\bar{q})$ being even
smaller due to OZI suppression\cite{cik}.
For $K\bar{K}$ molecules the rate was predicted
to be higher, $\sim (0.4 - 1) \times 10^{-4}$\cite{cik}, while for tightly
compact
$qq\bar{q}\bar{q}$ states the rate is yet higher, $\sim 2 \times
10^{-4}$\cite{achasov3,cik}.
Thus at first sight there seems to be a clear means to distinguish amongst
them.

In the $K\bar{K}$ molecule and $qq\bar{q}\bar{q}$ scenarios
it has uniformly been assumed that the radiative transition will be
driven by an intermediate  $K^+K^-$ loop ($\phi \to K^+K^-
\to \gamma K^+K^-  \to \gamma 0^{++}$). Explicit calculations
in the literature agree that this implies\cite{achasov3,cik,cbrown,
lucieu}

\begin{equation}
b.r. (\phi \to f_0(980)\gamma ) \sim 2 \pm 0.5 \times 10^{-4} \times F^2(R)
\end{equation}

\noindent where $F^2(R) = 1$ in point-like effective field theory
computations, such as refs.~\cite{achasov3,lucieu}. The range of
predicted magnitudes for the branching ratios in eq.~(1) reflect
the sensitivity to assumed parameters, such as masses and couplings
that vary slightly among these references.
By contrast, if the $f_0(980)$ and $ a_0(980)$ are spatially extended
$K\bar{K}$
molecules, (with r.m.s. radius $R > O(\Lambda_{QCD}^{-1}$)),
then  the high momentum region of the integration
in  refs.~\cite{cik,cbrown} is cut off, leading in effect
to a form factor suppression, $F^2(R) < 1$\cite{cik,achasov4}.
The differences in absolute rates are thus intimately linked to the
model dependent magnitude of $F^2(R)$.

In any event, one would expect in such pictures that
the branching ratio in eq.~(1) will not exceed $2.5 \times 10^{-4}$.
It is therefore tantalising that the measured rate\cite{PDG} (which
is quoted as an average of the data from refs.~\cite{cmd,snd}) appears to
be large:

\begin{equation}
b.r. (\phi \to  f_0 \gamma) = 3.4 \pm 0.4 \times 10^{-4}
\end{equation}

Therefore, precision data
on both $f_0$ and $a_0$ production, which are expected to
be forthcoming soon from DA$\Phi$NE,
will be most interesting.
Whatever the data may be, there are two particular items that we wish
to address concerning the current predictions.
One concerns the absolute branching ratios, and the second concerns the
ratio of branching ratios where,
if $f_0$ and $a_0^0$ have common constituents (and hence are ``siblings")
and are eigenstates of isospin, then their
affinity for $K^+K^-$ should be the same and so\cite{achasov3,cik,lucieu}

\begin{equation}
\frac{\Gamma(\phi \to  f_0 \gamma)}{\Gamma(\phi \to  a_0 \gamma)} \sim 1.
\end{equation}

 There are reasons to be suspicious of the predictions in both eqs. (1) and
(3).
We shall frame our remarks in the context of the $K\bar{K}$ molecule,
but they apply more generally.

If in the $K\bar{K}$ molecule one has

\begin{equation}
|f_0\rangle = cos\theta |K^+K^-\rangle + sin\theta |K^0\bar{K^0} \rangle
\end{equation}
\noindent and
\begin{equation}
|a_0^0\rangle = sin\theta |K^+K^-\rangle - cos\theta |K^0\bar{K^0} \rangle
\end{equation}

\noindent then the branching ratios $\phi \to \gamma f_0(\gamma a_0)$
 as found in ref.~\cite{cik}
can be summarised as follows

\begin{equation}
B.R.(\phi \to \gamma f_0:\gamma a_0) =  (4 \pm 1)
\times 10^{-4}(cos^2\theta:sin^2\theta)
(\frac{g^2_{SK^+K^-}/4\pi}{0.58GeV^2}) F^2(R)
\end{equation}

\noindent As shown in ref.~\cite{cik},
the analytical results of point-like effective field
theory calculations (e.g. refs.~\cite{achasov3,lucieu}) can be recovered
as $R \to 0$, for which
$F^2(R) \to 1$. In contrast to the compact hadronic
four quark state, the $K\bar{K}$ molecule is
spatially extended with r.m.s. $R \sim 1/\sqrt{m_K \epsilon}$, where $\epsilon$
is the binding energy and $F^2(R) < 1$,
 the precise magnitude depending on the $K\bar{K}$
molecular dynamics, which we shall discuss later.
It is clear also that the absolute rate in eq.~(6)
is driven by (i) the assumed
value for $\frac{g_{SK^+K^-}^2}{4\pi}  = 0.58$ GeV$^2$, and
(ii) the further assumption that the $f_0$ and $a_0$ are
$K\bar{K}$ states with $I=0,1$: hence
 $\theta = \pi/4$.

There are reasons to question both of these assumptions.

The assumed value $\frac{g_{f_0K^+K^-}^2}{4\pi}  = 0.58 GeV^2$ is consistent
with
that used in the effective field theory calculations of
refs.~\cite{achasov3,lucieu}. However, recent data raise some doubts as to
the reliability of this number, and it is not always clear in the
literature as to how this coupling is being defined.

We define the coupling of a scalar to two pseudoscalars, as follows.
For example, for the $f_0(980)$ which is above threshold for decay
into pions,

 $\Gamma(f \to \pi^+\pi^-) = \frac{g_{f\pi^+\pi^-}^2}{4\pi}\frac{1}{4m_f}
\sqrt{1-\frac{4m_{\pi}^2}{m_f^2}}$.

\noindent The determination of the
actual magnitude of the $g^2_{fK^+K^-}$ coupling
requires some care in view of
 the subtle ways that unitarity can affect the
$\pi \pi$ and $K\bar{K}$ couplings when the $K\bar{K}$ threshold is
opening, for which a  coupled channel analysis is required.

Recently
determinations of
the couplings of the $f_0$ to both $\pi\pi$ and to
$K\bar{K}$ have been measured in central production by
the WA102\cite{wa102gK} collaboration at CERN.
Their data are amenable to a coupled channel analysis and ref.~\cite{wa102gK}
found

 $\frac{g_{f\pi^+\pi^-}^2}{4\pi}  =
0.24 \pm 0.04 \pm 0.05 GeV^2 $

 $\frac{g_{fK^+K^-}^2}{4\pi}  =
0.39 \pm 0.04 \pm 0.04 GeV^2$

\noindent (Our convention related to that of refs.~\cite{wa102gK,e791} is
$\frac{g_{fK\bar{K}}^2}{4\pi} \equiv g_K \times 2m_f^2$ or
$\frac{g_{fK^+K^-}^2}{4\pi}(GeV)^2 \sim g_K $)

Thus, adopting this value, the predicted
rates would be correspondingly renormalised downwards by
$\frac{g_{fKK}^2}{4\pi}/0.58  = 0.67 \pm 0.10$
 which would make an even greater mismatch with the extant data.
Moreover, an analysis of Fermilab E791~\cite{e791}
data, which studies the $f_0(980)$ produced in $D_S$ decays, even suggests that
$\frac{g_{fK^+K^-}^2}{4\pi}  \sim 0.02 \pm 0.04 \pm 0.03$ (GeV)$^2$, hence
consistent with zero!
However, it should be noted that only the $\pi\pi$ decay mode of the $f_0(980)$
has been studied in this experiment and hence the coupling to $K^+K^-$ is
only measured indirectly.
With such uncertainties in the
value of this coupling strength, predictions of absolute rates for
$\phi \to \gamma f_0(980)$ or $\phi \to \gamma a_0(980)$ via an intermediate
$K\bar{K}$ loop must be treated with some caution.

By contrast, in the ratio of branching ratios this uncertainty is reduced, at
least
in the case of $K\bar{K}$ molecules for which\cite{cik}
$\frac{\Gamma(\phi \to  f_0 \gamma)}{\Gamma(\phi \to  a_0 \gamma)} \sim 1$.
Hence
a significant deviation from equality would appear to be a rather direct
elimination of $K\bar{K}$ molecules and, perhaps, other models where a
strong affinity of ``siblings" to the intermediate $K^+K^-$ state is assumed.
This also will
be important to test at DA$\Phi$NE as,
 within rather large errors, the results from ref.~\cite{snd}
in particular suggest that

\begin{equation}
\frac{\Gamma(\phi \to  f_0 \gamma)}{\Gamma(\phi \to  a_0 \gamma)} \sim
3.2 \pm 1.8
\end{equation}
in contrast to eq.~(3).

In this context, we draw attention to a potentially dramatic effect
upon the (relative and absolute)
 rates for
 $\phi \to
\gamma f_0(980)$ and $\phi \to
\gamma a_0(980)$ due to
 significant isospin mixing in the $f_0-a_0^0$
system\cite{ck1}.  This effect, which appears to be due to the
proximity to the $K\bar{K}$ threshold\cite{achasov1,speth}
and the differing mass gaps to the
$K^+K^-$ and $K^0\bar{K^0}$, could be amplified in $\phi$
radiative decays that proceed
{\it via} an intermediate $K\bar{K}$ loop\cite{achasov3,cik,cbrown,lucieu}.

Traditionally in strong interactions isospin has been
believed to be a nearly exact symmetry, broken only by
the slightly different masses of the $u$ and $d$ quarks and/or
electroweak effects. The small difference in mass between $K^{\pm}$
and $K^0$ is a particular example. However, the nearness of the
$\phi$ and
the $f_0(980)/a_0(980)$ to the $K^+K^-$ and $K^0K^0$ thresholds
causes the relative mass gaps to the charged and neutral
thresholds to be
 substantially different. As a result the dynamics of
such strongly coupled $K \bar{K}$ states \cite{iswein,jaffe2,tornqv,penn}
may be described better in a basis
specified by mass eigenstates. Such dynamics would give rise to
a violation of isospin and lead to mixing of states with
different G-parities.

\par
The possibility of such an effect was suggested long ago
in ref.~\cite{achasov1} and has been studied phenomenologically in
ref.~\cite{speth} and ref.~\cite{achasov2}.
The former, in particular,  has specifically drawn attention
to the relation between the existence of $K\bar{K}$ molecular bound
states and large violations of isospin.
These papers  all concentrated on the
production of the $f_0(980)/a_0(980)$ by flavoured mesons or photons;
in ref.~\cite{cik} we proposed that rather clean tests of the mixing
could be obtained from their production by gluonic systems,
such as the \pom (Pomeron)-induced
production in the central region at high energy:
$pp \to pp + f_0(980)/a_0(980)$.

Our analysis showed that new data from the WA102 collaboration
at CERN~\cite{etapipap}
are already consistent with a significant mixing. Specifically:
  in
(isoscalar) \pom (Pomeron)-induced
production in the central region at high energy,
production of the $a_0^0(980)$ comes dominantly from mixing with the
$f_0(980)$ such that the $f_0 - a_0$ are not good isospin eigenstates.
In the language of the $K\bar{K}$ molecule, at least, this
would translate into  $\theta \neq \frac{\pi}{4}$ in eq.~(6)
and hence to a
 significant difference in behaviour for $\Gamma(\phi \to \gamma f_0)/
\Gamma(\phi \to \gamma a_0)$.

With the basis as defined in eqs.~(4) and (5),
the ratio of production rates by \pom \pom (isoscalar)
fusion in central production will be

\begin{equation}
\sigma (\pome \pome \to a_0)/\sigma (\pome \pome \to f_0)
 = \frac{1-sin 2\theta}{1+sin 2 \theta}
\end{equation}

\noindent In ref.~\cite{ck1} we found this to be $ (8 \pm 3) \times 10^{-2}$.
Hence if we assume that the production phase is the same for the two,
we obtain

\begin{equation}
cot \theta = 1.8 \pm 0.2  ( \theta = 30^o \pm 3^o)
\end{equation}
\noindent and hence predict that within this approximation
the relative rates will be

\begin{equation}
\frac{\Gamma(\phi \to \gamma f_0)}{\Gamma(\phi \to \gamma a_0)} \equiv
cot^2\theta
= 3.2 \pm 0.8
\end{equation}

\noindent This is far from the naive expectation of unity for ideal isospin
states and
in remarkable agreement with data (eq.~(7)).

This prediction, eq.~(10), is
 a rather direct consequence of the isospin mixing obtained
in ref.~\cite{ck1}. In order to use the data to abstract magnitudes
of $F^2(R)$, and hence assess how compact the four-quark
state is,a definitive accurate value for $g_{fKK}^2/4\pi $ will be required.

If for orientation we adhere to the value used elsewhere, $g_{fKK}^2/4\pi\sim
0.6$
GeV$^2$,
 and impose the preferred $\theta$,
then the results of ref.~\cite{cik} are revised to

\begin{equation}
b.r.(\phi \to \gamma f_0) + b.r.(\phi \to \gamma a_0) \leq (4 \pm 1)
\times 10^{-4}
\end{equation}

\noindent and

\begin{equation}
b.r.(\phi \to \gamma f_0) =  (3.0 \pm 0.6)  \times 10^{-4} F^2(R)
\end{equation}
\begin{equation}
b.r.(\phi \to \gamma a_0) =  (1.0 \pm 0.25) \times 10^{-4} F^2(R)
\end{equation}

 For illustration we cite two models.  Barnes\cite{barnes}
developed a simple potential picture of a
$K\bar{K}$ molecule,
ignoring any short range annihilation and
rescattering corrections.  This leads to a high momentum
cut-off in the $K^+K^-$ loop. Following Barnes' parameterisation,
ref.~\cite{cik} described the high momentum cut off by a power law, such that
the $K^+K^- 0^{++}$ vertex form factor $\phi(p) =  \mu^4/(p^2 + \mu^2)^2$
in which case
$\mu \equiv \sqrt{3}/2R$. This led to
$R \sim 1.2 fm$, $F^2(R) \sim 0.25$.

 However, the predictions are rather sensitive to the assumed details.
For example, the authors of ref.~\cite{cik} also considered
a Gaussian parameterisation for the
$K^+K^- 0^{++}$ vertex form factor $\phi(p) =  e^{-p^2/4\mu_0}$ and
$\mu_0 \equiv 3/16R^2$.
Barnes'
parameters in this case imply that $\mu_0 \sim 0.4 fm^{-2}$ and $R \sim 0.7
fm$.
in which case
the suppression is only some 20\%; $F^2(R) \sim 0.8$.
In more sophisticated treatments, the role of
annihilation involving non-$K\bar{K}$
intermediate states such as $\pi \pi$ and $\pi \eta$ will modify the
potential.

If experiment confirms the predicted ratio in eq.~(10),
then the individual rates
may be used as a measure of $F^2(R)$. Branching ratios for
which $F^2(R) << 1$ would imply that the $K^+K^-0^{++}$
interaction is spatially extended, $R > O(\Lambda_{QCD}^{-1})$. Conversely,
if $F^2(R) \to 1$, the system is spatially compact, as in $qq\bar{q}\bar{q}$.
If,
as preliminary data suggest, the rates are between these extremes, then
a qualitative picture may emerge of a compact structure, such as $q\bar{q}$
 or $qq\bar{q}\bar{q}$,
which spends a sizeable part of its lifetime in a two meson state, such
as $K\bar{K}$. Such a picture has also recently been suggested, based on
QCD sum rules for an intrinsic $s\bar{s}$ ``seed", in ref.~\cite{fazio}.


\begin{center}
{\bf Acknowledgements}
\end{center}
\par
This work is supported, in part, by grants from
the British Particle Physics and Astronomy Research Council
and the European Community Human Mobility Program Eurodafne,
contract NCT98-0169.

\clearpage

\end{document}